\documentclass[12pt,a4paper]{article}

\usepackage[colorinlistoftodos]{todonotes}

\usepackage{ifthen} 
\usepackage{booktabs}
\newboolean{pdflatex}
\setboolean{pdflatex}{true} 

\newboolean{articletitles}
\setboolean{articletitles}{true} 

\newboolean{uprightparticles}
\setboolean{uprightparticles}{false} 

\newboolean{inbibliography}
\setboolean{inbibliography}{false} 

\def\paperauthors{LHCb collaboration} 
\def\paperasciititle{Search for the lepton-flavour violating decays BToKem} 
\def\papertitle{Search for the lepton-flavour violating decays \BToKem} 
\def\paperkeywords{{High Energy Physics}, {LHCb}} 
\def\papercopyright{\the\year\ CERN for the benefit of the LHCb collaboration} 
\def\paperlicence{CC-BY-4.0 licence}
\def\paperlicenceurl{https://creativecommons.org/licenses/by/4.0/}

\usepackage[top=1in, bottom=1.25in, left=1in, right=1in]{geometry}

\usepackage{microtype}
\usepackage{lineno}  
\usepackage{xspace} 
\usepackage{caption} 

\usepackage{graphicx}  
\usepackage{color}
\usepackage{colortbl}
\graphicspath{{./figs/}} 
\DeclareGraphicsExtensions{.pdf,.PDF,png,.PNG}

\usepackage{amsmath} 
\usepackage{amssymb}
\usepackage{amsfonts}
\usepackage{upgreek} 

\newcommand*\patchAmsMathEnvironmentForLineno[1]{%
\expandafter\let\csname old#1\expandafter\endcsname\csname #1\endcsname
\expandafter\let\csname oldend#1\expandafter\endcsname\csname
end#1\endcsname
 \renewenvironment{#1}%
   {\linenomath\csname old#1\endcsname}%
   {\csname oldend#1\endcsname\endlinenomath}%
}
\newcommand*\patchBothAmsMathEnvironmentsForLineno[1]{%
  \patchAmsMathEnvironmentForLineno{#1}%
  \patchAmsMathEnvironmentForLineno{#1*}%
}
\AtBeginDocument{%
\patchBothAmsMathEnvironmentsForLineno{equation}%
\patchBothAmsMathEnvironmentsForLineno{align}%
\patchBothAmsMathEnvironmentsForLineno{flalign}%
\patchBothAmsMathEnvironmentsForLineno{alignat}%
\patchBothAmsMathEnvironmentsForLineno{gather}%
\patchBothAmsMathEnvironmentsForLineno{multline}%
\patchBothAmsMathEnvironmentsForLineno{eqnarray}%
}

\usepackage{hyperxmp}

\usepackage[pdftex,
            pdfauthor={\paperauthors},
            pdftitle={\paperasciititle},
            pdfkeywords={\paperkeywords},
            pdfcopyright={Copyright (C) \papercopyright},
            pdflicenseurl={\paperlicenceurl}]{hyperref}

\usepackage[all]{hypcap} 

\usepackage{xspace} 
\usepackage{upgreek}

\def\BToKem     {\decay{\Bp}{\Kp \mupm \emp}}

\def\BToKmSS     {\decay{\Bp}{\Kp \mup \en}}
\def\BToKeSS     {\decay{\Bp}{\Kp \mun \ep}}
\def\BToJpsiK  {\decay{\Bp}{\Kp \jpsi ( \to \mup \mun)}}

\def\BToJpsieeK  {\decay{\Bp}{\Kp \jpsi ( \to \ep \en)}}

\def\lhcb   {\mbox{LHCb}\xspace}

\def\babar  {\mbox{BaBar}\xspace}

\def\MagUp {\mbox{\em Mag\kern -0.05em Up}\xspace}

\ifthenelse{\boolean{uprightparticles}}%
{

 \def\Pmu         {\ensuremath{\upmu}\xspace}                 
 \def\Pnu         {\ensuremath{\upnu}\xspace}                 
                  
 \def\Ppi         {\ensuremath{\uppi}\xspace}

 \def\Ppsi        {\ensuremath{\uppsi}\xspace}

 \def\PDelta      {\ensuremath{\Delta}\xspace}                 
 \def\PXi      {\ensuremath{\Xi}\xspace}                 
 \def\PLambda      {\ensuremath{\Lambda}\xspace}                 
 \def\PSigma      {\ensuremath{\Sigma}\xspace}                 
 \def\POmega      {\ensuremath{\Omega}\xspace}                 
 \def\PUpsilon      {\ensuremath{\Upsilon}\xspace}                 
 
 \def\PB      {\ensuremath{\mathrm{B}}\xspace}                 
                  
 \def\PD      {\ensuremath{\mathrm{D}}\xspace}

 \def\PJ      {\ensuremath{\mathrm{J}}\xspace}                 
 \def\PK      {\ensuremath{\mathrm{K}}\xspace}

 \def\Pb      {\ensuremath{\mathrm{b}}\xspace}

 \def\Pe      {\ensuremath{\mathrm{e}}\xspace}

 \def\Pi      {\ensuremath{\mathrm{i}}\xspace}

 \def\Ps      {\ensuremath{\mathrm{s}}\xspace}

}
{

 \def\Pmu         {\ensuremath{\mu}\xspace}                 
 \def\Pnu         {\ensuremath{\nu}\xspace}                 
                  
 \def\Ppi         {\ensuremath{\pi}\xspace}

 \def\Ppsi        {\ensuremath{\psi}\xspace}                 
                  
 \mathchardef\PDelta="7101
 \mathchardef\PXi="7104
 \mathchardef\PLambda="7103
 \mathchardef\PSigma="7106
 \mathchardef\POmega="710A
 \mathchardef\PUpsilon="7107
                  
 \def\PB      {\ensuremath{B}\xspace}                 
                  
 \def\PD      {\ensuremath{D}\xspace}

 \def\PJ      {\ensuremath{J}\xspace}                 
 \def\PK      {\ensuremath{K}\xspace}

 \def\Pb      {\ensuremath{b}\xspace}

 \def\Pe      {\ensuremath{e}\xspace}

 \def\Pi      {\ensuremath{i}\xspace}

 \def\Ps      {\ensuremath{s}\xspace}

}

\makeatletter
\ifcase \@ptsize \relax
  \newcommand{\miniscule}{\@setfontsize\miniscule{4}{5}}
\or
  \newcommand{\miniscule}{\@setfontsize\miniscule{5}{6}}
\or
  \newcommand{\miniscule}{\@setfontsize\miniscule{5}{6}}
\fi
\makeatother

\DeclareRobustCommand{\optbar}[1]{\shortstack{{\miniscule (\rule[.5ex]{1.25em}{.18mm})}
  \\ [-.7ex] $#1$}}

\def\en         {{\ensuremath{\Pe^-}}\xspace}   
\def\ep         {{\ensuremath{\Pe^+}}\xspace}
\def\epm        {{\ensuremath{\Pe^\pm}}\xspace} 
\def\emp        {{\ensuremath{\Pe^\mp}}\xspace}

\def\mup        {{\ensuremath{\Pmu^+}}\xspace}
\def\mun        {{\ensuremath{\Pmu^-}}\xspace} 
\def\mupm       {{\ensuremath{\Pmu^\pm}}\xspace} 
\def\mump       {{\ensuremath{\Pmu^\mp}}\xspace}

\def\neu        {{\ensuremath{\Pnu}}\xspace}
\def\neub       {{\ensuremath{\overline{\Pnu}}}\xspace}
\def\neul       {{\ensuremath{\neu_\ell}}\xspace}
\def\neulb      {{\ensuremath{\neub_\ell}}\xspace}

\def\squark    {{\ensuremath{\Ps}}\xspace}

\def\bquark    {{\ensuremath{\Pb}}\xspace}

\def\pion   {{\ensuremath{\Ppi}}\xspace}

\def\pip    {{\ensuremath{\pion^+}}\xspace}
\def\pim    {{\ensuremath{\pion^-}}\xspace}

\def\kaon    {{\ensuremath{\PK}}\xspace}
  \def\Kbar    {{\kern 0.2em\overline{\kern -0.2em \PK}{}}\xspace}

\def\KorKbar    {\kern 0.18em\optbar{\kern -0.18em K}{}\xspace}

\def\Kp      {{\ensuremath{\kaon^+}}\xspace}
\def\Km      {{\ensuremath{\kaon^-}}\xspace}

  \def\Dbar    {{\kern 0.2em\overline{\kern -0.2em \PD}{}}\xspace}
\def\D       {{\ensuremath{\PD}}\xspace}

\def\DorDbar    {\kern 0.18em\optbar{\kern -0.18em D}{}\xspace}
\def\Dz      {{\ensuremath{\D^0}}\xspace}
\def\Dzb     {{\ensuremath{\Dbar{}^0}}\xspace}

\def\Dstarp  {{\ensuremath{\D^{*+}}}\xspace}

\def\B       {{\ensuremath{\PB}}\xspace}
\def\Bbar    {{\ensuremath{\kern 0.18em\overline{\kern -0.18em \PB}{}}}\xspace}

\def\BorBbar    {\kern 0.18em\optbar{\kern -0.18em B}{}\xspace}

\def\Bu      {{\ensuremath{\B^+}}\xspace}

\def\Bp      {{\ensuremath{\Bu}}\xspace}

\def\jpsi     {{\ensuremath{{\PJ\mskip -3mu/\mskip -2mu\Ppsi\mskip 2mu}}}\xspace}

  \def\Y#1S{\ensuremath{\PUpsilon{(#1S)}}\xspace}

\def\Lz          {{\ensuremath{\PLambda}}\xspace}
\def\Lbar        {{\ensuremath{\kern 0.1em\overline{\kern -0.1em\PLambda}}}\xspace}
\def\LorLbar    {\kern 0.18em\optbar{\kern -0.18em \PLambda}{}\xspace}

\def\Lb      {{\ensuremath{\Lz^0_\bquark}}\xspace}

\newcommand{\decay}[2]{\mbox{\ensuremath{#1\!\to #2}}\xspace}         

\def\to                 {\ensuremath{\rightarrow}\xspace}

\def\CP                {{\ensuremath{C\!P}}\xspace}

\def\AT#1     {\ensuremath{A_{\mathrm{T}}^{#1}}\xspace}           

\def\C#1      {\ensuremath{\mathcal{C}_{#1}}\xspace}                       
\def\Cp#1     {\ensuremath{\mathcal{C}_{#1}^{'}}\xspace}                    
\def\Ceff#1   {\ensuremath{\mathcal{C}_{#1}^{\mathrm{(eff)}}}\xspace}        
\def\Cpeff#1  {\ensuremath{\mathcal{C}_{#1}^{'\mathrm{(eff)}}}\xspace}       
\def\Ope#1    {\ensuremath{\mathcal{O}_{#1}}\xspace}                       
\def\Opep#1   {\ensuremath{\mathcal{O}_{#1}^{'}}\xspace}                    

\newcommand{\tev}{\ifthenelse{\boolean{inbibliography}}{\ensuremath{~T\kern -0.05em eV}}{\ensuremath{\mathrm{\,Te\kern -0.1em V}}}\xspace}
\newcommand{\gev}{\ensuremath{\mathrm{\,Ge\kern -0.1em V}}\xspace}
\newcommand{\mev}{\ensuremath{\mathrm{\,Me\kern -0.1em V}}\xspace}
\newcommand{\kev}{\ensuremath{\mathrm{\,ke\kern -0.1em V}}\xspace}
\newcommand{\ev}{\ensuremath{\mathrm{\,e\kern -0.1em V}}\xspace}
\newcommand{\gevc}{\ensuremath{{\mathrm{\,Ge\kern -0.1em V\!/}c}}\xspace}
\newcommand{\mevc}{\ensuremath{{\mathrm{\,Me\kern -0.1em V\!/}c}}\xspace}
\newcommand{\gevcc}{\ensuremath{{\mathrm{\,Ge\kern -0.1em V\!/}c^2}}\xspace}
\newcommand{\gevgevcccc}{\ensuremath{{\mathrm{\,Ge\kern -0.1em V^2\!/}c^4}}\xspace}
\newcommand{\mevcc}{\ensuremath{{\mathrm{\,Me\kern -0.1em V\!/}c^2}}\xspace}

\def\invfb   {\ensuremath{\mbox{\,fb}^{-1}}\xspace}

\def\gsim{{~\raise.15em\hbox{$>$}\kern-.85em
          \lower.35em\hbox{$\sim$}~}\xspace}
\def\lsim{{~\raise.15em\hbox{$<$}\kern-.85em
          \lower.35em\hbox{$\sim$}~}\xspace}

\def\pt         {\ensuremath{p_{\mathrm{ T}}}\xspace}

\def\evtgen     {\mbox{\textsc{EvtGen}}\xspace}

\def\geant      {\mbox{\textsc{Geant4}}\xspace}

\def\photos     {\mbox{\textsc{Photos}}\xspace}

\def\pythia     {\mbox{\textsc{Pythia}}\xspace}

\def\tell1  {TELL1\xspace}
\def\ukl1   {UKL1\xspace}

\newcommand{\eg}{\mbox{\itshape e.g.}\xspace}

\newcommand{\cf}{\mbox{\itshape cf.}\xspace}

\usepackage{cite} 
\usepackage{mciteplus}

\usepackage{longtable} 

\begin{document}

\renewcommand{\thefootnote}{\fnsymbol{footnote}}
\setcounter{footnote}{1}

\begin{titlepage}
\pagenumbering{roman}

\vspace*{-1.5cm}
\centerline{\large EUROPEAN ORGANIZATION FOR NUCLEAR RESEARCH (CERN)}
\vspace*{1.5cm}
\noindent
\begin{tabular*}{\linewidth}{lc@{\extracolsep{\fill}}r@{\extracolsep{0pt}}}
\ifthenelse{\boolean{pdflatex}}
{\vspace*{-1.5cm}\mbox{\!\!\!\includegraphics[width=.14\textwidth]{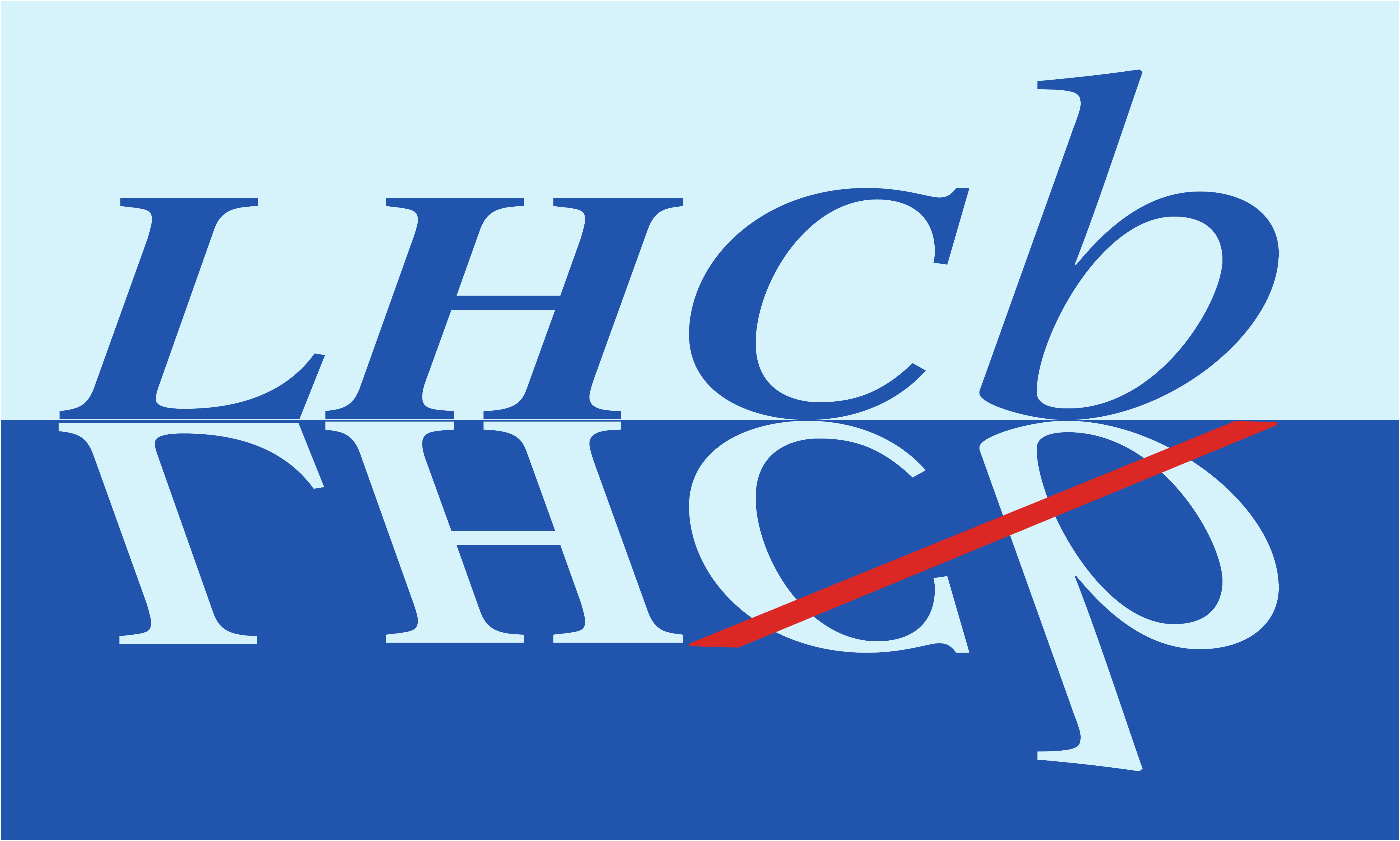}} & &}%
{\vspace*{-1.2cm}\mbox{\!\!\!\includegraphics[width=.12\textwidth]{lhcb-logo.eps}} & &}%
\\
 & & CERN-EP-2019-172 \\  
 & & LHCb-PAPER-2019-022 \\  
 & & September 3, 2019\\
 & & \\
\end{tabular*}

\vspace*{4.0cm}

{\normalfont\bfseries\boldmath\huge
\begin{center}
  \papertitle 
\end{center}
}

\vspace*{2.0cm}

\begin{center}

\paperauthors\footnote{Authors are listed at the end of this paper.}
\end{center}

\vspace{\fill}

\begin{abstract}
  \noindent
   A search  for  the  lepton-flavour violating  decays \BToKem is performed using a sample of proton-proton collision data, collected with the \lhcb experiment at centre-of-mass energies of $7$  and $8\tev$  and corresponding to an integrated luminosity of 3\invfb. No significant signal is observed, and upper limits on the branching fractions are set as  \mbox{$\mathcal{B}(\BToKeSS) < 7.0~(9.5) \times 10^{-9}$} and \mbox{$\mathcal{B}(\BToKmSS) < 6.4~(8.8) \times 10^{-9}$} at 90 (95)\% confidence level. The results improve the current best limits on these decays by more than one order of magnitude.
\end{abstract}

\vspace*{2.0cm}

\begin{center}
  Submitted to
  Phys.~Rev.~Lett. 
\end{center}

\vspace{\fill}

{\footnotesize 
\centerline{\copyright~\papercopyright. \href{\paperlicenceurl}{\paperlicence}.}}
\vspace*{2mm}

\end{titlepage}


\newpage
\setcounter{page}{2}
\mbox{~}

\cleardoublepage

\renewcommand{\thefootnote}{\arabic{footnote}}
\setcounter{footnote}{0}

\pagestyle{plain} 
\setcounter{page}{1}
\pagenumbering{arabic}



The observation of neutrino oscillations has provided the first evidence for lepton-flavour violation (LFV) in neutral leptons. By contrast, LFV in the charged sector is negligible in the Standard Model (SM)~\cite{Raidal:2008jk} and any observation of a charged LFV decay would be indisputable evidence for physics beyond the SM (BSM). In light of recent flavour anomalies in semileptonic \decay{\bquark}{\squark\ell^{+}\ell^{-}} transitions~\cite{LHCb-PAPER-2019-009, LHCb-PAPER-2017-013,LHCb-PAPER-2015-051}, many SM extensions have been proposed that link lepton-universality violation to LFV, predicting in particular a significantly enhanced decay rate in \decay{\bquark}{\squark \mump \epm} processes. In this Letter a search for the decays of \BToKem is reported (Inclusion of charge-conjugate processes is implied throughout the letter). Their branching fractions are predicted to be in the range $10^{-8} - 10^{-10}$ in leptoquark models~\cite{Varzielas:2015iva,Hiller:2016kry}, extended gauge boson models~\cite{Crivellin:2015era}, or models including \CP violation in the neutrino sector~\cite{Boucenna:2015raa}. Currently, the best limits of $\mathcal{B}(\BToKeSS) < 9.1 \times 10^{-8}$ and $\mathcal{B}(\BToKmSS) < 13 \times 10^{-8}$  have been set by the \babar collaboration at the $90\%$ confidence level~\cite{Aubert:2006vb}.

A data set of proton-proton ($pp$) collisions corresponding to an integrated luminosity of 3\invfb, recorded with the \lhcb detector in 2011 and 2012 at centre-of-mass energies of $7\tev$ and $8\tev$, respectively, is used in this analysis. The two final states with different lepton charge combinations are studied independently, since they could be affected differently by BSM dynamics. The yields of the \BToKem decays are normalised to those of the \BToJpsiK decay, which has a well-known branching fraction~\cite{PDG2018}, the same topology, and similar signatures in the detector. The \BToJpsieeK decay is also used as a control channel in the analysis.


The \lhcb detector is a single-arm forward spectrometer covering the pseudorapidity
range $2 < \eta < 5$, described in detail in Refs.~\cite{Alves:2008zz,LHCb-DP-2014-002}.
The detector includes a silicon-strip
vertex detector surrounding the $pp$ interaction region, tracking stations located either side of a dipole magnet, ring-imaging Cherenkov (RICH) detectors, calorimeters and muon chambers.

The online event selection is performed by a trigger~\cite{LHCb-DP-2012-004},
which consists of a hardware stage, based on information from the calorimeter and muon
systems, followed by a software stage, which applies a full event
reconstruction. At the hardware trigger stage, \BToKem and \BToJpsiK event candidates are required to have a muon with high transverse momentum, \pt. In the subsequent software trigger, at least one charged particle must have a $\pt > 1.7\gevc$ in the 2011 data set and $\pt > 1.6 \gevc$ in 2012, unless the particle is identified as a muon in which case $\pt>1.0\gevc$ is required. This track must be significantly displaced from any primary interaction vertex (PV) in the event. Finally, a two- or three-track secondary vertex with a significant displacement from any PV is required, where a multivariate algorithm~\cite{BBDT} is used for the identification of secondary vertices consistent with the weak decay of a \bquark hadron.

Simulated samples are used to evaluate signal efficiencies, to train multivariate classifiers, to model the shape of the invariant mass of the \BToKem signal candidates, and to study physics backgrounds. In the simulation, $pp$ collisions are generated using
\pythia~\cite{Sjostrand:2007gs,Sjostrand:2006za} with a specific \lhcb
configuration~\cite{LHCb-PROC-2010-056}.  Decays of unstable particles
are described by \evtgen~\cite{Lange:2001uf}, in which final-state
radiation is generated using \photos~\cite{Golonka:2005pn}. A phase-space model is adopted for signal \BToKem decays.
The interaction of the generated particles with the detector, and its response,
are implemented using the \geant
toolkit~\cite{Allison:2006ve, *Agostinelli:2002hh} as described in
Ref.~\cite{LHCb-PROC-2011-006}.


The \BToKem candidates passing the trigger selection are reconstructed by combining three charged tracks originating from a good-quality common vertex. The tracks forming the \Bp candidate are required not to originate from any PV and must have sizeable transverse momentum. Due to the long lifetime of the \Bp meson, this vertex is required to be well separated from any PV. The \Bp direction vector, determined from its production and decay vertex positions, must be aligned with its momentum vector. The mass of the reconstructed \Bp candidate, $m(K^+\mu^\pm e^\mp)$, is restricted to lie within $\pm1500 \mevcc$ of the known \Bp meson mass~\cite{PDG2018}. Furthermore, the \B-meson decay products must be well identified as a kaon, an electron and a muon, exploiting information from the Cherenkov detectors, the calorimeters, and the muon stations. The electron candidate kinematics are corrected for bremsstrahlung photon emission if a compatible photon candidate in the calorimeter is found. Kaon and electron candidates that have hits in the muon stations consistent with their trajectories are rejected. The same selection is applied to the normalisation (control) channels, for which the dimuon (dielectron) invariant mass is additionally required to be consistent with the known \jpsi mass~\cite{PDG2018}. The selection and analysis procedures were developed without inspecting the signal data in the region $m(K^+\mu^\pm e^\mp) \in  \left[4985 , 5385 \right]\mevcc$.

The most significant backgrounds originate from partially reconstructed \Bp decays, \eg from double semileptonic \decay{\Bp}{\Dzb X \ell^+\neul} with \decay{\Dzb}{\Kp Y \ell^{\prime-}\neub_{\ell^\prime}} decays, where $X$ and $Y$ represent hadrons, while $\ell$ and $\ell^{\prime}$ are leptons. They are removed by imposing the requirement $m(K^+ \ell^-) > 1885 \mevcc$. Contributions from decays involving  charmonium  resonances, where one lepton is misidentified as a kaon or as a lepton of a different flavour, are rejected by mass vetoes.

The combinatorial background, which consists of  random  tracks that are  associated to a common vertex, is reduced using a boosted decision tree (BDT)~\cite{Breiman,AdaBoost} algorithm. This BDT combines information about the \Bp meson kinematics and information related to its flight distance, decay vertex quality and impact parameter with respect to the primary vertex. In addition, it uses information such as the impact parameters of the electron, muon and kaon candidates, and the isolation of the \Bp candidate from any other charged track in the event~\cite{Aaltonen:2007ad}.

The BDT is trained on simulated \BToKem events that have satisfied the previous requirements. The simulated samples are adjusted using \BToJpsiK and \BToJpsieeK decays in data to correct data-simulation differences in the \B-meson production kinematics, vertex quality, and detector occupancy represented by the number of tracks in the detector. The upper-mass sideband, corresponding to $m(K^+\mu^\pm e^\mp) \in [5385, 6000] \mevcc$, is used as a proxy for the background. The training is performed using a $k$-folding approach~\cite{Bevan:2017stx} with ten folds, which allows the whole sample to be used without biasing the output of the classifier.  The optimal requirement on the BDT classifier is chosen to give the best expected upper limits on the branching fractions $\mathcal{B}(\BToKem)$.

The candidates surviving  this multivariate selection are used to train a second BDT, dedicated to reject background from partially reconstructed $b$-hadron decays. The background sample for the training is taken from the lower-mass sideband in data, corresponding to $m(K^+\mu^\pm e^\mp) \in \left[4550, 4985\right] \mevcc$, where the partially reconstructed background is expected to contribute. The signal proxy is the same as for the first BDT. The training procedure shares the $k$-folding approach and the same set of discriminating variables used to construct the first multivariate discriminant, with the addition of the ratio between the projections of the electron and the $K^+\mu^\pm$ momenta orthogonal to the \B-meson direction of flight.
The requirements on the second BDT are optimized in the same manner as the first BDT. 
The final stage of the selection, where requirements on the particle identification (PID) variables based on a neural net classifier for the kaon, electron and muon are applied~\cite{Aaij:2308409}, ensures the suppression of candidates from decays with misidentification of at least one particle.

The performance of the PID algorithms is not perfectly simulated, and thus a correction is performed using high-purity calibration data samples of muons from \decay{\B}{X\jpsi(\to\mu^+\mu^-)} decays, electrons from \BToJpsieeK decays, and kaons from \decay{\Dstarp}{\Dz(\to \Km\pip ) \pip} decays~\cite{LHCb-PUB-2016-021}. The calibration data are binned in the particle's momentum and pseudorapidity, and in the detector occupancy. Particle identification variables for the simulated data sets are sampled from the distributions of calibration data in the corresponding bin. The performance of the PID resampling is validated on both the \BToJpsiK and \BToJpsieeK control channels.

The potential contamination from \bquark-hadron decays in the signal mass region after selection is analysed using dedicated simulated samples. Two categories are analysed: fully reconstructed $B$ decays, with at least one particle in the final state misidentified, such as the semileptonic decays  \decay{\Bp}{\Kp \ell^+\ell^-} and \decay{\Bp}{\Kp \jpsi(\to \ell^+\ell^-) }, or  fully hadronic \Bp decays as \decay{\Bp}{\Kp\pip\pim}; partially reconstructed decays in which at least one particle is not reconstructed and one or more particles are misidentified in addition, such as  \decay{B^0}{K^{*0} \ell^+\ell^-},   \decay{\Lb}{p \Km \ell^+\ell^-}, \decay{\Lb}{p \Km \jpsi(\to \ell^+\ell^-)} and  \decay{\Bp}{\Dzb \ell^+ \neul} transitions, where the \Dzb meson decays further
to $\Kp \pim$ or $\Kp \ell^- \neulb$. 
The expected number of candidates from each possible background source after the selection is evaluated from simulation and is found to be negligible.


The branching fraction $\mathcal{B}(\BToKem)$ is measured relative to the normalisation channel using
\begin{align}
\label{eqn:BRalpha}
\mathcal{B}(\BToKem) & =   N(\BToKem)  \times \alpha, \\ \alpha& \equiv \frac{ { \mathcal{B}(\BToJpsiK) }}{\varepsilon(\BToKem)} \frac{\varepsilon(\BToJpsiK)}{N(\BToJpsiK)}, \nonumber
\end{align}
where the $\varepsilon(\BToJpsiK)$ and $\varepsilon(\BToKem)$ denote the efficiencies of the normalisation and signal channels, respectively; $N( \BToJpsiK)$ and $N(\BToKem)$ are the observed \BToJpsiK and \BToKem yields, respectively. The value of the branching fraction of the normalisation mode is $ \mathcal{B}(\BToJpsiK) = (6.02 \pm 0.17) \times 10^{-5}$, taken from Ref.~\cite{PDG2018}. The yield of the normalisation channel is determined from an unbinned extended maximum-likelihood
 fit to the invariant mass $m(K^+\mu^+\mu^-)$ of the  selected \BToJpsiK candidates, performed separately on 2011 and 2012 data.  The sum of two Crystal Ball functions~\cite{Skwarnicki:1986xj} is  used to parameterise the signal, while an exponential function models the background. The yields resulting from the fits are $26940 \pm 170$ for 2011 and $59220 \pm 250$ for 2012 data.

\begin{figure}[tbp]
\centering
\includegraphics[width=0.45\linewidth]{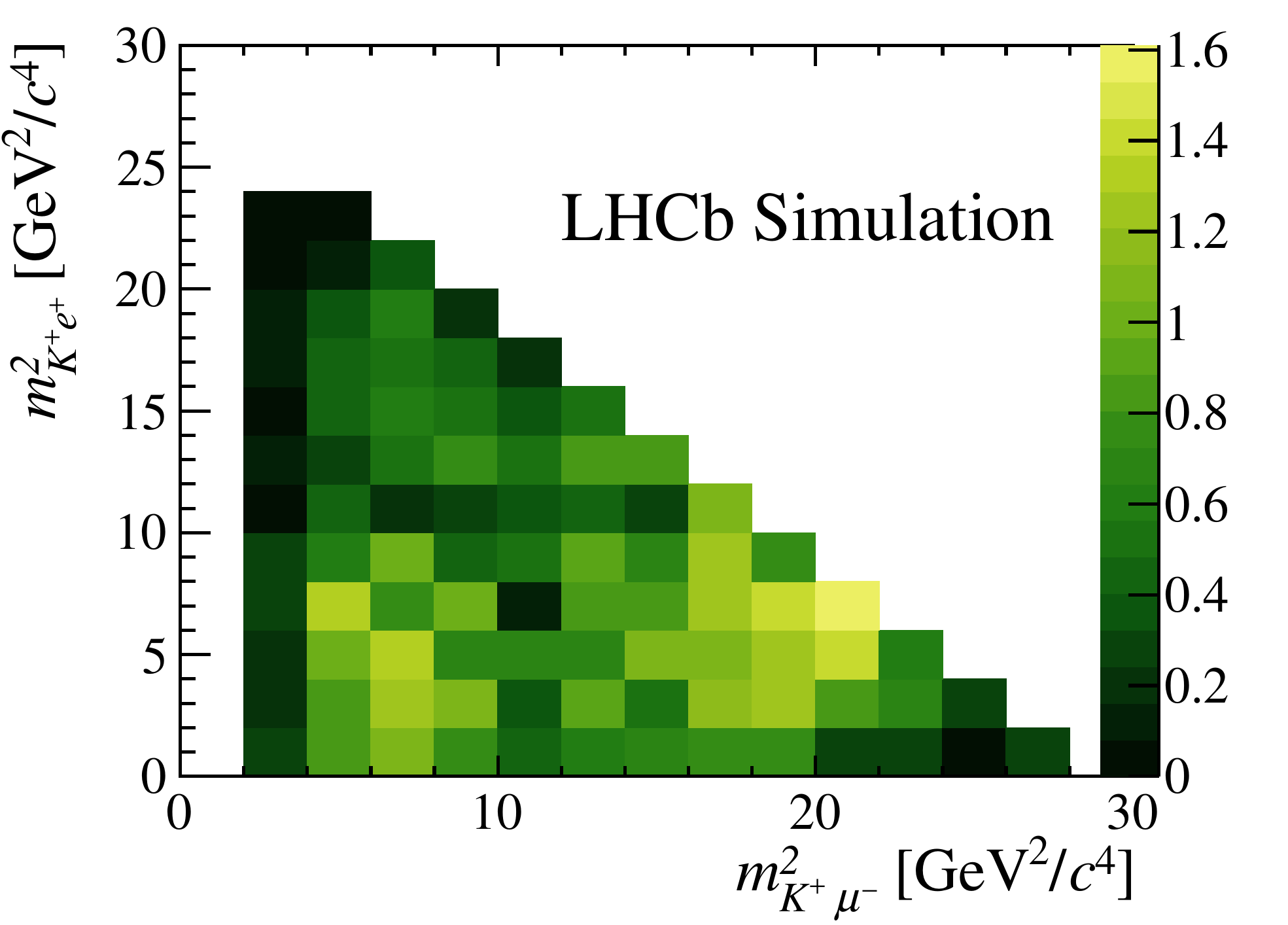}
\includegraphics[width=0.45\linewidth]{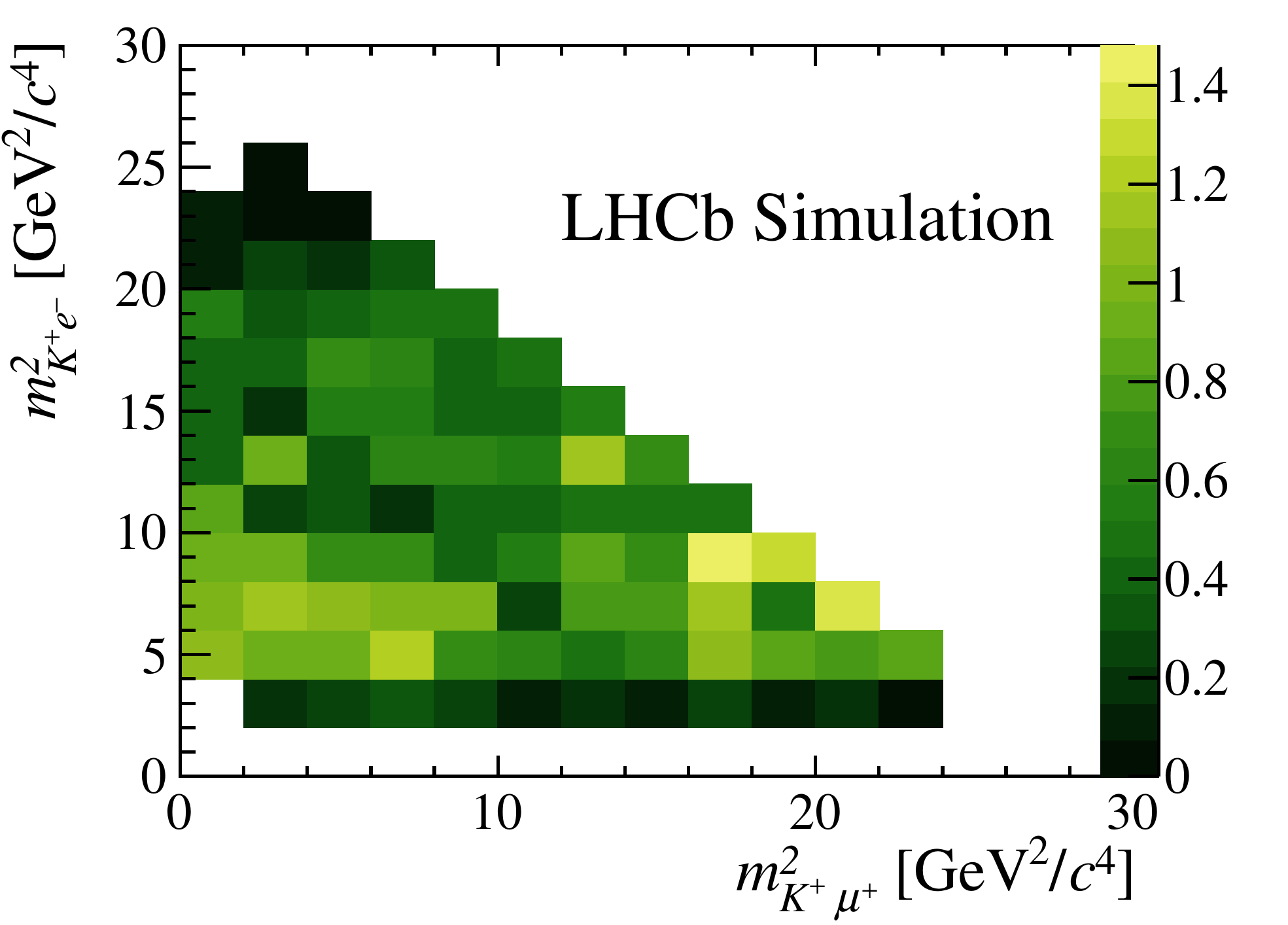}
 \caption{Efficiency of (left) \BToKeSS and (right) \BToKmSS as function of the squared invariant masses $m^2_{K^+e^\pm}$ and $m^2_{K^+\mu^\mp}$. The variation of efficiency across the Dalitz plane is due to applied vetoes. The efficiencies are given in per mille.}
 \label{fig:dalitzEff}
 \end{figure}

The efficiencies are calculated taking into account all selection requirements. The analysis is performed assuming a phase-space model for the signal decay. Efficiency maps in bins of the invariant masses of the particles in the final state $m^2_{K^+e^\pm}$ and $m^2_{K^+\mu^\mp}$ are provided in Fig~\ref{fig:dalitzEff} to allow for the interpretation of the result in different BSM scenarios.

All efficiencies are determined from calibrated simulation samples and the normalisation factors for the two decay channels are given in Table~\ref{tab:alpha}. The two data taking periods are combined into  a single normalisation factor taking into account the relative data sizes and efficiencies.
The ratio $\alpha/\mathcal{B}(\BToJpsiK)$, which excludes external inputs, is also quoted.

\begin{table}[tbp]
	\centering
	\caption{Normalisation factor $\alpha$ for \BToKeSS and \BToKmSS  final states. The ratio $\alpha/\mathcal{B}(\BToJpsiK)$ is independent of external inputs.}
	\begin{tabular}{lccc}
 		Decay  & $\alpha$&$\alpha/\mathcal{B}(\BToJpsiK)$\\
		\toprule
		\BToKeSS 	&  $(1.97 \pm 0.14)\times10^{-9}$& $(3.27 \pm 0.22)\times10^{-5}$\\
		\BToKmSS 	&  $(2.21 \pm 0.14)\times10^{-9}$& $(3.68 \pm 0.22)\times10^{-5}$\\
		\end{tabular}
	\label{tab:alpha}
\end{table}


The invariant-mass distribution of  \BToKem candidates  is modeled differently depending on whether bremsstrahlung photons have been included in the momentum calculation for the electrons. The sum of two Crystal Ball functions with common mean value is used in both cases. In the case of bremsstrahlung the tails are on opposite sides of the peak. Otherwise, the two tails share the same parameters. Their values, obtained from the \BToKem simulation, are corrected taking into account the differences between data and simulation in  \BToJpsiK and \BToJpsieeK decays.  Two types of unbinned maximum-likelihood fits are performed on the dataset. The first fit assumes only background is present, with the background modeled with an exponential function. From this fit $3.9 \pm 1.1$ and $0.9 \pm 0.6$ background candidates are expected in the signal mass window for the \BToKmSS and  \BToKeSS modes, respectively. The second fit includes the signal component, which is used to determine the signal yields. The  \BToKeSS and \BToKmSS invariant-mass distributions are fitted separately.


\begin{figure}[tbp]
\centering
\includegraphics[width=0.45\linewidth]{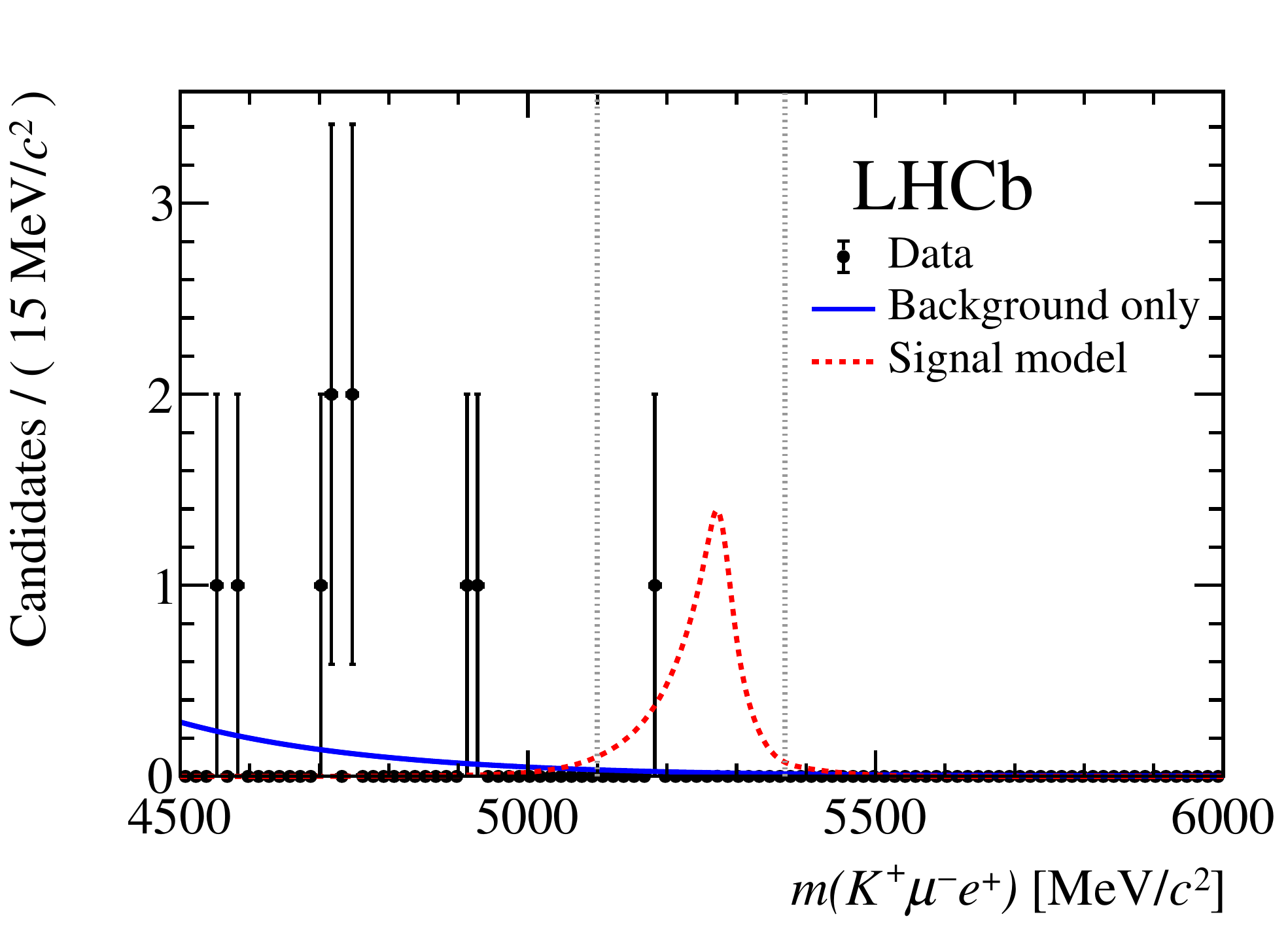}
\includegraphics[width=0.45\linewidth]{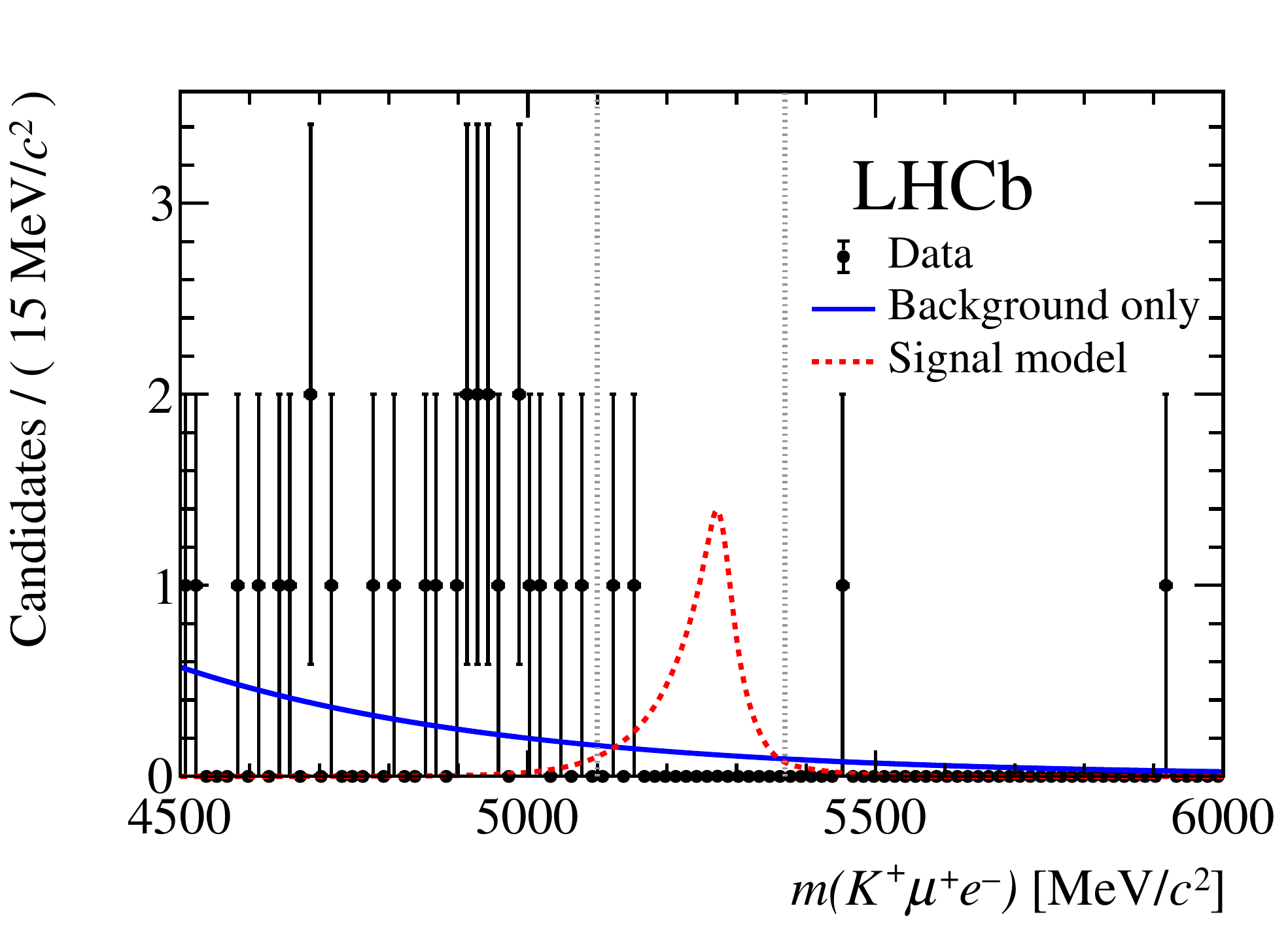}
 \caption{Invariant-mass distributions of the (left) \BToKeSS and (right) \BToKmSS candidates obtained on the combined data sets recorded in 2011 and 2012 with background only fit functions (blue continuous line) and the signal model normalised to 10 candidates (red dashed line) superimposed. The signal window is indicated with grey dotted lines. Difference between the two distributions arises from the effect of the $m(K^+\ell^-)$ requirement.}
  \label{fig:datafit}
 \end{figure}

\begin{figure}[tbp]
\centering
\includegraphics[width=0.45\linewidth]{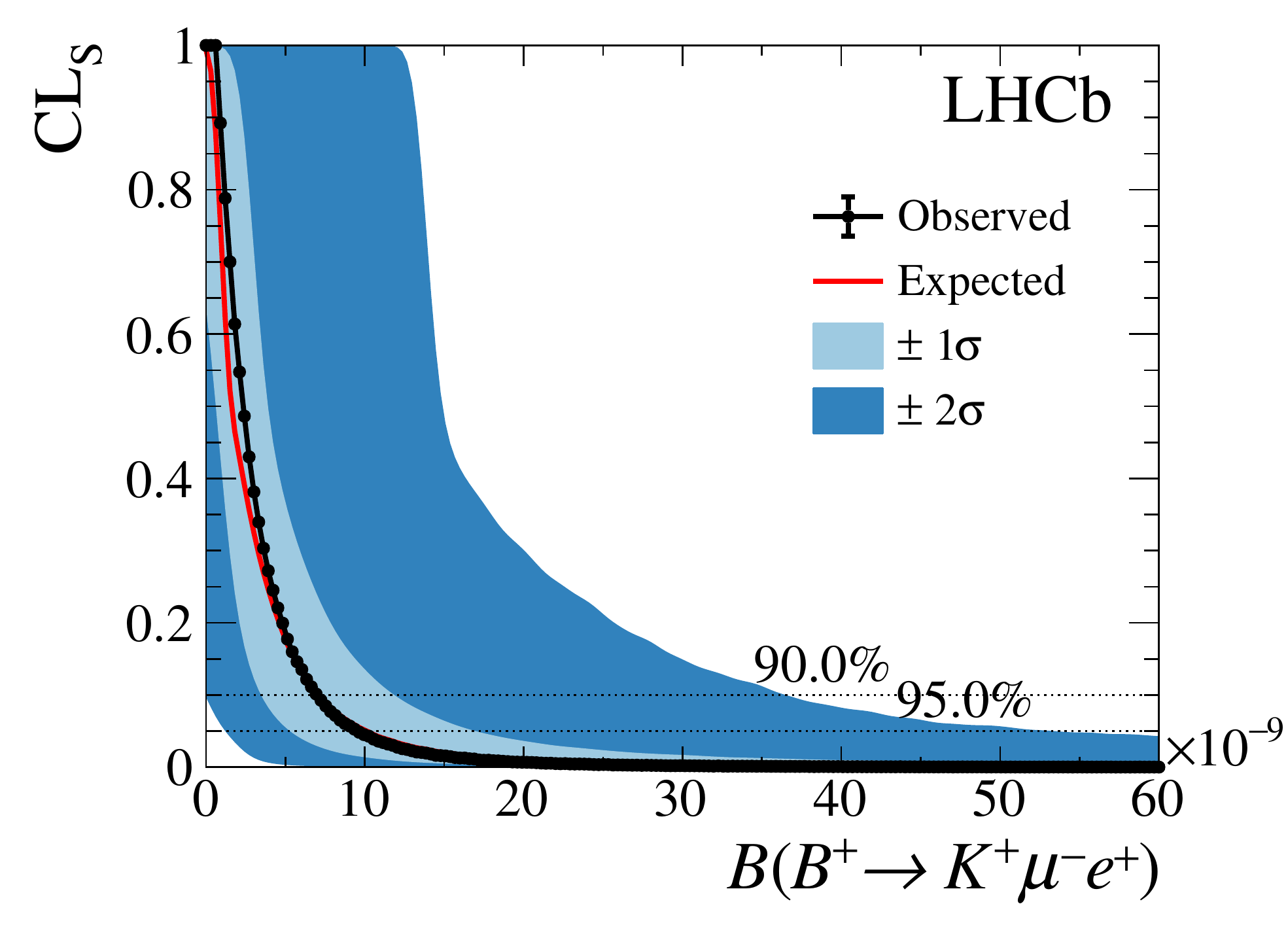}
\includegraphics[width=0.45\linewidth]{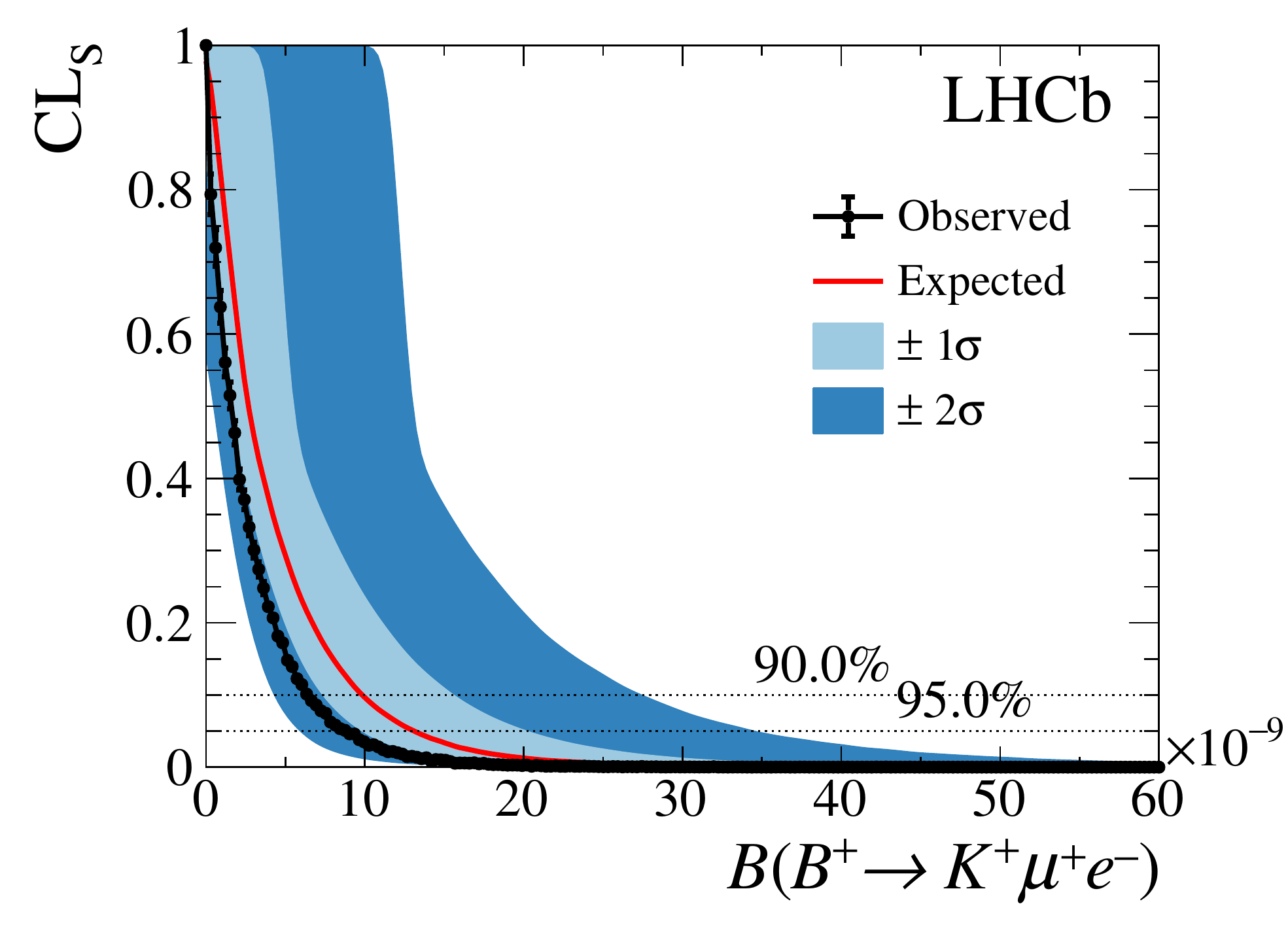}
 \caption{Upper limits on the branching fractions of (left) \BToKeSS and (right) \BToKmSS decays obtained on the combined data sets recorded in 2011 and 2012. The red solid line (black solid line with data points) corresponds to the distribution of the expected (observed) upper limits, and the light blue (dark blue) band contains the $1\sigma$ ($2\sigma$) uncertainties.}
 \label{fig:limits}
 \end{figure}

After unblinding the data set, there are $1$ $(2)$ candidates in the signal mass window $m(K^+\mu^\pm e^\mp) \in \left[5100 , 5370 \right]~\mevcc$ for the \BToKeSS (\BToKmSS) channels, respectively, in agreement with the background-only hypothesis (\cf Fig.~\ref{fig:datafit}). 
The upper limits on the branching fractions are set with the CL${_s}$ method~\cite{CLs}, using  the GammaCombo framework~\cite{Aaij:2016kjh,matthew_kenzie_2019_3373613} with a one-sided test statistic. The likelihoods are computed from fits to the invariant-mass distributions with the normalisation constant constrained to its nominal value accounting for statistical and systematic uncertainties. Pseudoexperiments, in which the nuisance parameters are input at their best fit value and the background yield is varied according to its systematic uncertainty, are used for the evaluation of the test statistic.
The resulting upper limits are shown in Fig.~\ref{fig:limits} and Table~\ref{tab:res}.

\begin{table}[tbp]
	\centering
	\caption{Upper limit on the branching fraction of \BToKeSS and \BToKmSS decays obtained on the combined data sets recorded in 2011 and 2012 for confidence levels of $90\%$ and $95\%$.}
	\begin{tabular}{lcc}
		 &  {$90\%$ C. L.} & {$95\%$ C. L.}\\
		\toprule
		$\mathcal{B}$(\BToKeSS)$/ 10^{-9}$ 	& 7.0 & 9.5\\
		$\mathcal{B}$(\BToKmSS)$/ 10^{-9}$ 	& 6.4 & 8.8 \\
		\end{tabular}
	\label{tab:res}
\end{table}

The dominant sources of systematic uncertainty on the upper limits are due to applied simulation corrections. These include the kinematic difference of \B-meson production, residual difference between correcting the muon and electron candidates, and PID resampling. Furthermore, the determination of trigger efficiencies and the knowledge of the background invariant-mass distribution are also considered in evaluating the systematic uncertainty.

The systematic uncertainty on the sampling procedure of the PID variables includes two components. The first stems from applying the {\textit{sPlot}}~\cite{Pivk:2004ty} method to the calibration data, and adds an uncertainty of $0.1\%$  for kaons and muons, and $3\%$ for electrons,  the latter being a conservative estimate originating from a comparison of the {\textit{sPlot}} method with a cut-and-count method. The second component addresses the choice of binning in the sampling procedure. It is evaluated by recalculating the normalisation factor $\alpha$ using a finer and a coarser binning, and taking the largest deviation with respect to the baseline result.

A small difference in the correction procedure is observed depending on the choice of control channel, namely \BToJpsiK or \BToJpsieeK. This difference, referred to as electron-muon difference, is taken as systematic uncertainty. 

The systematic uncertainty from the fitting model is determined to be $2.1\%$ using a bootstrapping approach. The systematic uncertainty on the background model is calculated by repeating the fit using an alternative model, where the exponential function is obtained from a sample enriched in background events. The difference between the alternative and nominal background parametrization is taken as a systematic uncertainty. The uncertainty on the knowledge of the \BToJpsiK branching fraction is combined with the uncertainty due to the limited size of the simulation sample and is propagated to the normalisation constant, corresponding to a systematic uncertainty of $3.5\%$.
A summary of systematic uncertainties is reported in Tab.~\ref{tab:syst}.

\begin{table}[tb]
	\centering
	\caption{Summary of systematic uncertainties.}
	\begin{tabular}{ccc}
		{Effect} &  \BToKmSS & \BToKeSS \\
		\toprule
                Data-simulation corrections & {$1.0\%$} & {$1.0\%$} \\
                Electron-muon differences & {$1.4\%$} & {$1.4$\%} \\
                Fitting model & {$2.1\%$} & {$2.1\%$} \\
                PID resampling & {$4.5\%$} & {$5.5\%$} \\
                Trigger  & {$1.0\%$} & {$1.0\%$} \\
                Normalisation factor & {$3.5\%$} & {$3.5\%$} \\
    \midrule
    Total & {$6.4\%$} & {$7.1\%$} \\
    \midrule
                Background & {$0.60$} & {$0.43$} \\
		\bottomrule
	\end{tabular}
	\label{tab:syst}
\end{table}
%


In conclusion, a search for the lepton-flavour violating decays \BToKem is performed using data from proton-proton collisions recorded with the \lhcb experiment at centre-of-mass energies of 7\tev and 8\tev, corresponding to an integrated luminosity of 3\invfb. A uniform distribution of signal events within the phase space accessible to the final-state particles is assumed. No excess is observed over the background-only hypothesis, and the resulting upper limits on the branching fractions are \mbox{$\mathcal{B}(\BToKeSS) <  7.0~(9.5) \times 10^{-9}$} and \mbox{$\mathcal{B}(\BToKmSS) < 6.4~(8.8) \times 10^{-9}$}  at  $90\, (95)\%$ confidence level. The results improve the current best limits on the decays~\cite{Aubert:2006vb} by more than one order of magnitude. Moreover, the measurements impose strong constraints on the aforementioned extensions to the SM~\cite{Varzielas:2015iva,Hiller:2016kry,Boucenna:2015raa,Crivellin:2015era}.


\section*{Acknowledgements}
%
%
\noindent We express our gratitude to our colleagues in the CERN
accelerator departments for the excellent performance of the LHC. We
thank the technical and administrative staff at the LHCb
institutes.
We acknowledge support from CERN and from the national agencies:
CAPES, CNPq, FAPERJ and FINEP (Brazil); 
MOST and NSFC (China); 
CNRS/IN2P3 (France); 
BMBF, DFG and MPG (Germany); 
INFN (Italy); 
NWO (Netherlands); 
MNiSW and NCN (Poland); 
MEN/IFA (Romania); 
MSHE (Russia); 
MinECo (Spain); 
SNSF and SER (Switzerland); 
NASU (Ukraine); 
STFC (United Kingdom); 
DOE NP and NSF (USA).
We acknowledge the computing resources that are provided by CERN, IN2P3
(France), KIT and DESY (Germany), INFN (Italy), SURF (Netherlands),
PIC (Spain), GridPP (United Kingdom), RRCKI and Yandex
LLC (Russia), CSCS (Switzerland), IFIN-HH (Romania), CBPF (Brazil),
PL-GRID (Poland) and OSC (USA).
We are indebted to the communities behind the multiple open-source
software packages on which we depend.
Individual groups or members have received support from
AvH Foundation (Germany);
EPLANET, Marie Sk\l{}odowska-Curie Actions and ERC (European Union);
ANR, Labex P2IO and OCEVU, and R\'{e}gion Auvergne-Rh\^{o}ne-Alpes (France);
Key Research Program of Frontier Sciences of CAS, CAS PIFI, and the Thousand Talents Program (China);
RFBR, RSF and Yandex LLC (Russia);
GVA, XuntaGal and GENCAT (Spain);
the Royal Society
and the Leverhulme Trust (United Kingdom).


\appendix

\clearpage

\addcontentsline{toc}{section}{References}
\setboolean{inbibliography}{true}
\bibliographystyle{LHCb}
\bibliography{main,standard,LHCb-PAPER,LHCb-CONF,LHCb-DP,LHCb-TDR}

\newpage
\centerline{\large\bf LHCb collaboration}
\begin{flushleft}
\small
R.~Aaij$^{28}$,
C.~Abell{\'a}n~Beteta$^{46}$,
T.~Ackernley$^{56}$,
B.~Adeva$^{43}$,
M.~Adinolfi$^{50}$,
C.A.~Aidala$^{77}$,
Z.~Ajaltouni$^{6}$,
S.~Akar$^{61}$,
P.~Albicocco$^{19}$,
J.~Albrecht$^{11}$
% ...

{\footnotesize \it
% ...
$ ^{6}$Universit{\'e} Clermont Auvergne, CNRS/IN2P3, LPC, Clermont-Ferrand, France\\
% ...
$ ^{11}$Fakult{\"a}t Physik, Technische Universit{\"a}t Dortmund, Dortmund, Germany\\
% ...
$ ^{19}$INFN Laboratori Nazionali di Frascati, Frascati, Italy\\
% ...
$ ^{28}$Nikhef National Institute for Subatomic Physics, Amsterdam, Netherlands\\
% ...
$ ^{43}$Instituto Galego de F{\'\i}sica de Altas Enerx{\'\i}as (IGFAE), Universidade de Santiago de Compostela, Santiago de Compostela, Spain\\
% ...
$ ^{46}$Physik-Institut, Universit{\"a}t Z{\"u}rich, Z{\"u}rich, Switzerland\\
% ...
$ ^{50}$H.H. Wills Physics Laboratory, University of Bristol, Bristol, United Kingdom\\
% ...
$ ^{56}$Oliver Lodge Laboratory, University of Liverpool, Liverpool, United Kingdom\\
% ...
$ ^{61}$University of Cincinnati, Cincinnati, OH, United States\\
% ...
$ ^{63}$Syracuse University, Syracuse, NY, United States\\
% ...
$ ^{77}$University of Michigan, Ann Arbor, United States, associated to $^{63}$\\
% ...
}
\end{flushleft}

\end{document}